\documentclass[aps,twocolumn,pra,superscriptaddress,showpacs,tightenlines]{revtex4}
\usepackage{mathrsfs}
\usepackage{amssymb}
\usepackage{amsmath}
\usepackage{graphicx}
\usepackage{epsfig}
\usepackage{txfonts}
\usepackage{subfigure}
\usepackage{amsfonts}

\begin{document}

\title{Quantum correlations of two qubits interacting with a macroscopic medium}

\author{Yan Liu}
\affiliation{Key Laboratory of Low-Dimensional Quantum Structures
and Quantum Control of Ministry of Education, and Department of
Physics, Hunan Normal University, Changsha 410081, China}
\author{Jing Lu}
\thanks{Corresponding author}
\email{lujing@hunnu.edu.cn} \affiliation{Key Laboratory of
Low-Dimensional Quantum Structures and Quantum Control of Ministry
of Education, and Department of Physics, Hunan Normal University,
Changsha 410081, China}
\author{Lan Zhou}
\affiliation{Key Laboratory of Low-Dimensional Quantum Structures and Quantum Control of Ministry
of Education, and Department of Physics, Hunan Normal University,
Changsha 410081, China}

\date{\today}

\begin{abstract}
We consider two particles of spin-$1/2$ interacting with a
one-dimensional $N$-spin array, which is an exactly solvable model.
The dynamics of entanglement and quantum discord (QD) of the spins
of the two particles is investigated by regarding the 1D N-spin
array as the environment. It is found that although the entanglement
may suffers a sudden death and a sudden birth in the evolution, it
can neither be generated nor become larger than its initial value.
Different from the entanglement dynamics, QD can be amplified, and
even be generated by the interaction between particles and the
common environment. We also observe that QD decays asymptotically to
zero and later experiences a rival when the average number of
excitation in the 1D N-spin array becomes larger in the case of
nonzero inter-distance between two particles.
\end{abstract}


\maketitle \narrowtext


\section{Introduction}

To find the fundamental resource for quantum information processing tasks,
quantum entanglement has been regarded as a necessary prerequisite in order
for quantum correlations to occur. With the doubting on entanglement being
responsible for all quantum speedups raised by Knill and Laflamme \cite%
{KLPRL81(98)}, it is discovered that superposition principle can entail an
alternative type of nonclassical correlation apart from entanglement. There
are several measure to characterize all nonclassical correlations \cite%
{RMP84(12)1655}. Quantum discord \cite{QDZerekPRL88,VedralJPA34} is the
prominent one, which is defined as the difference between the quantum
generalizations of two classically equivalent versions of the mutual
information. Quantum discord is original introduced to be an
information-theoretic approach to decoherence mechanisms in a quantum
measurement process.

A quantum measurement process typically involves a measured system
and a measuring apparatus, where the measured system is described as
a quantum system and the apparatus behaves as a classical object.
And the process of measurement can be described as a dynamic
evolution process via an appropriate coupling between a measured
system and the measuring apparatus. To draw light on the border
between the quantum and the classical, the macroscopic character of
the apparatus is taken into account by Hepp and Coleman. In their
exactly solvable model, known as the Coleman-Hepp (CH) model
\cite{HCmodel}, the apparatus is a \textquotedblleft\ large
system\textquotedblright\ and the number of its constituting
components approaches infinity, where each component obeys the
Schr\"{o}dinger equation and interactes with the measured system
locally. Later on, a modified version of CH model is proposed by
Nakazato and Pascazio \cite{HCmodel-1}, which takes energy exchange
between the measured system and the apparatus into account. Based on
exact solvability of the Coleman-Hepp model and its generalized
version \cite{HCmodel-1,HCmodel-2,HCmodel-3}, the emergence of
classicality of a quantum system is due to the elimination of the
off-diagonal elements of the density matrix, which is characterized
by a factorization structure \cite{HCmodel-2} due to the interaction
between the quantum system and the classical object. And this
factorization gives rise to the exponential behavior of a quantum
system \cite{HCmodel-4}. However, to take the information transfer
associated with the measured system and the apparatus into account,
a triple model is necessary for a quantum measurement process, which
consists of a measured system, an apparatus, and an environment.
This invokes the investigation of the behavior of correlations of a
two-qubit composite system exposed to noisy environments
\cite{IndeEnv-1,IndeEnv-2,IndeEnv-3,IndeEnv-4,IndeEnv-5,IndeEnv-6,comEnv-1,EPL106,CSBZhang,QIC
Yao,QIC Guo}. However, many of these studies concern on independent
environments, and approximations are usually employed.

Exactly solvable models, which exclude the effects of
approximations, not only gives us good insights into physics, but
they also provide us with a better understanding of the complicated
phenomena involved, for example, the CH model and its generalized
version \cite{HCmodel-1,HCmodel-2,HCmodel-3} have given us a
comprehensive understanding of the quantum measurement processes. In
this paper, we study two particles with spin-$1/2$ interacting with
a one-dimensional (1D) N-spin array, which is a generalized version
of CH model. We regarded the spins of two particles and the 1D
$N$-spin array as a two-qubit composite system and its surrounding
environment, respectively. The 1D $N$-spin array is a macroscopic
system as $N$ becomes larger. To introduce the macroscopic medium of
the 1D array, we first study the effect of the macroscopic system on
one qubit, it is found that the dynamic evolution of the qubit can
be modeled as a phase damping channel in the weak-coupling
macroscopic limit. Next, quantum correlations of the spins of two
particles are investigated. It is found that quantum discord can be
amplified, even generated via the interaction. We also observed the
revival of quantum discord and entanglement under some conditions.

This paper is organized as follows. In Sec.~\ref{Sec:2}, we introduce the
model---two particles with spin-$1/2$ interacting with 1D array made of $N$
identical spins. In Sec.~\ref{Sec:3}, we derive the time evolution operator
of the total system, and also present the damping effect of the 1D array on
the spin state of one particle. In Sec.~\ref{Sec:4}, we study quantum
correlation of the two-qubit in the time evolution. Finally, We conclude
this work in Sec.~\ref{Sec:5}.


\section{\label{Sec:2}the model}

The CH model describes an ultrarelativistic particle interacting with 1D
array of $N$ identical spins. Each spin of the 1D array could be regarded as
a AgBr molecule with the down (up) state corresponding to the undivided
(dissociated) molecule \cite{HCmodel}. In the present generalized version,
we consider two particles $Q_{A}$ and $Q_{B}$. Each particle possesses a
spin, and moves in the $x$ direction with a positive constant velocity $v_{j}
$ ($j=A,B$). The 1D $N$-spin array has the finite size with length $%
L=x_{N}-x_{1}$, where $x_{1}$ ($x_{N}$) is the position of the first (final)
spin of the array. After the particles arriving at position $x_{1}>0$, the
particles begin their interaction with the spin array.

The Hamiltonian of the total system
\begin{equation}
H=H_{Q}+H_{E}+H^{\prime }
\end{equation}%
is a sum of three parts, where
\begin{equation}
H_{Q}=v_{A}\hat{p}_{A}+v_{B}\hat{p}_{B}+\sum_{j=A,B}\frac{1}{2}\hbar \omega
_{j}\left( 1+\tau _{j}^{z}\right)  \label{2eq-01}
\end{equation}%
is the free Hamiltonian of two particles,%
\begin{equation}
H_{E}=\frac{1}{2}\hbar \omega \sum_{n=1}^{N}\left( 1+\sigma _{n}^{z}\right)
\label{2eq-02}
\end{equation}%
is the free Hamiltonian of 1D $N$-spin array, and
\begin{eqnarray}
H^{\prime } &=&\sum_{j=A,B}\frac{1+\tau _{j}^{z}}{2}H_{jE}^{\prime }
\label{2eq-03} \\
&\equiv &\sum_{j=A,B}\sum_{n=1}^{N}\frac{1+\tau _{j}^{z}}{2}V\left( \hat{x}%
_{j}-x_{n}\right) \left( \sigma _{n}^{+}e^{-i\frac{\omega }{v_{j}}\hat{x}%
_{j}}+h.c.\right)  \notag
\end{eqnarray}%
is the interaction Hamiltonian between the particles and the 1D $N$-spin
array. Here, $\sigma _{n}^{\alpha }$ ($\alpha =\pm ,z$) is the Pauli spin
operators for the $n$th spin of the 1D array, $\tau _{j}^{z}$ is the Pauli
matrice acting on the spin of the $j$th particle. $\hat{x}_{j}\ $and $\hat{p}%
_{j}$ are the coordinator and momentum operator of the $Q_{j}$ particle,
which satisfies the canonical commutation relation $\left[ \hat{x}_{j},\hat{p%
}_{j^{\prime }}\right] =i\hbar \delta _{jj^{\prime }}$. The real function $%
V\left( \hat{x}_{j}-x_{n}\right) $ characterizes the strength of the
interaction between the particles and each spin of the 1D array, where $x_{n}
$ $(n=1,...,N)$ are the positions of the spin in the array ($x_{n}>x_{n-1}$%
). Hereafter, the subscript $n$ denotes the spatial location of each spin in
the array.

Comparing to the original CH model, the interaction Hamiltonian is changed
to taking the spins of the particles into account, which describes that the $%
j$th particle at the position $x$ exerts a real potential $V\left( \hat{x}%
_{j}-x_{n}\right) $ on the $n$th spin if and only if particle's spin
is up \cite{HCmodel-3}. Denoting $\left\vert p_{j}\right\rangle $ as
the state that the $j$th particle has momentum $p$ and $\left\vert
0_{n}\right\rangle $ ($\left\vert 1_{n}\right\rangle $) as the down
(up) state of the $n$th spin, Hamiltonian $H_{jn}^{\prime }$ changes
the states as
\begin{subequations}
\begin{eqnarray}
H_{jn}^{\prime }\left\vert p_{j}\right\rangle \left\vert
0_{n}\right\rangle
&=&V\left( \hat{x}_{j}-x_{n}\right) \left\vert p_{j}-\frac{\hbar \omega }{%
v_{j}}\right\rangle \left\vert 1_{n}\right\rangle ,  \label{2eq-04} \\
H_{jn}^{\prime }\left\vert p_{j}\right\rangle \left\vert 1_{n}\right\rangle
&=&V\left( \hat{x}_{j}-x_{n}\right) \left\vert p_{j}+\frac{\hbar \omega }{%
v_{j}}\right\rangle \left\vert 0_{n}\right\rangle ,
\end{eqnarray}%
\end{subequations} where Hamiltonian
\begin{equation}
H_{jn}^{\prime }=V\left( \hat{x}_{j}-x_{n}\right) \left( \sigma _{n}^{+}e^{-i%
\frac{\omega }{v_{j}}\hat{x}_{j}}+h.c.\right)  \label{2eq-05}
\end{equation}%
is the $H_{jE}^{\prime }$ term acting on the $n$th spin and the $j$th
particle. Consequently, once the particles' spin is up, the interaction
Hamiltonian $H_{jE}^{\prime }$ given by Eq.~(\ref{2eq-03}) induces the
energy exchange between the particles and the spin array due to the
nonvanishing energy gap $\hbar \omega $ between the two states of the
molecule \cite{HCmodel-1}. The particle's spin undergoes free precession
with frequency $\omega _{j}$.

For later convenience, the Hamiltonian of the total system is rearranged as
the sum of the free Hamiltonian $H_{0}$ and the interaction Hamiltonian $%
H^{\prime }$ of the particle-array system
\begin{equation}
H=H_{0}+H^{\prime },  \label{2eq-06}
\end{equation}%
where $H_{0}=H_{Q}+H_{E}$.


\section{\label{Sec:3} The dynamics of the total system}

For a given initial state of the total system $\rho \left( 0\right) $, the
state at time $t$ reads $\rho \left( t\right) =U\left( t\right) \rho \left(
0\right) U^{\dag }\left( t\right) $. Obviously, the evolution operator $%
U\left( t\right) =e^{-iHt/\hbar }$ characterizes the dynamical of
the particle-array system, where the initial time is set as
$t_{0}=0$. To exactly solve the present model, we introduce the
interaction picture by writing the evolution operator as $U\left(
t\right) =e^{-iH_{0}t/\hbar }U_{AB}$, where the unitary operator
$U_{AB}\left( t\right) $ satisfies the Schr\"{o}dinger equation with
the Hamiltonian (\ref{2eq-03}). Since the total spin of the
particles $Q_{A}$ and $Q_{B}$ along the $z$ direction is always
conserved during the time evolution, we decompose the unitary
operator as
\begin{eqnarray}
U_{AB}\left( t\right) &=&U_{A}\left( t\right) U_{B}\left( t\right)
\left\vert \uparrow \uparrow \right\rangle \left\langle \uparrow \uparrow
\right\vert +\left\vert \downarrow \downarrow \right\rangle \left\langle
\downarrow \downarrow \right\vert +  \label{3eq-01} \\
&&U_{B}\left( t\right) \left\vert \downarrow \uparrow \right\rangle
\left\langle \downarrow \uparrow \right\vert +U_{A}\left( t\right)
\left\vert \uparrow \downarrow \right\rangle \left\langle \uparrow
\downarrow \right\vert ,  \notag
\end{eqnarray}%
where states $\left\vert \downarrow \right\rangle $ and $\left\vert \uparrow
\right\rangle $ are the eigenstates of the operator $\tau ^{z}$. Here, the
reduced evolution operators $U_{j}\left( t\right) $ obey the following
Schr\"{o}dinger equation%
\begin{equation}
i\hbar \partial _{t}U_{j}\left( t\right) =H_{jE}^{\prime }U_{j}\left(
t\right) ,j=A,B,  \label{3eq-02}
\end{equation}%
which can be computed exactly as%
\begin{align}
U_{j}\left( x_{j},t\right) & =\exp \left[ -\frac{i}{\hbar }%
\sum_{n=1}^{N}\int_{0}^{t}dt^{\prime }V\left( x_{j}+v_{j}t^{\prime
}-x_{n}\right) \right.  \label{3eq-03} \\
& \times \left. \left( \sigma _{n}^{+}e^{-i\frac{\omega }{v_{j}}x_{j}}
+\sigma _{n}^{-}e^{i\frac{\omega }{v_{j}}x_{j}}\right) \right]
\notag
\end{align}%
in the coordinator representation. With the SU(2) algebra, the
exponential (10) can be disentangled as
\begin{equation}
U_{j}\left( x_{j},t\right) =\prod\limits_{n}e^{-i\sigma _{n}^{+}\tan \alpha
_{n}^{\left[ j\right] }\left( t\right) }e^{-\sigma _{n}^{z}\ln \cos \alpha
_{n}^{\left[ j\right] }\left( t\right) }e^{-i\sigma _{n}^{-}\tan \alpha
_{n}^{\left[ j\right] }\left( t\right) },  \label{3eq-04}
\end{equation}%
where we have defined the time-dependent tipping angles \cite{HCmodel-4} of
the $n$th spin as
\begin{equation}
\alpha _{n}^{\left[ j\right] }\left( t\right) =\int_{0}^{t}\frac{dt^{\prime }%
}{\hbar }V\left( x_{j}+v_{j}t^{\prime }-x_{n}\right) .  \label{3eq-05}
\end{equation}%
For the sake of simplicity, we restrict our attention to the case of $\delta
$-shaped potentials, i.e., assuming that $V\left( x\right) =V_{0}\Omega
\delta \left( x\right) $. It allows us to obtain the tipping angles exactly
as
\begin{equation}
\alpha _{n}^{\left[ j\right] }\left( t\right) =\frac{V_{0}\Omega }{\hbar
v_{j}}\Theta \left( x_{j}+v_{j}t-x_{n}\right) ,  \label{3eq-06}
\end{equation}%
where $\Theta \left( y\right) $ is the Heaviside unit step function, i.e., $%
\Theta \left( y\right) =1$ for $y>0$, and $\Theta \left( y\right) =0$ for $%
y<0$. In Eq.~(\ref{3eq-06}), we have assumed that the spin array is placed
at the far right of the origin ($x_{1}>0$), and the two particles are
initially at the position $x_{A}$ and $x_{B}$ with $x_{A},x_{B}< x_{1}$,
i.e., well outside the potential region of the spin array. It can observed
that if two particles, initially at the same position $x_{A}=x_{B}=x$, move
with the same constant velocities, i.e., $v_{j}=v$, tipping angles of the $n$%
th spin are equal, $\alpha _{n}^{\left[ j\right] }\left( t\right) =\alpha
_{n}\left( t\right) =\frac{V_{0}\Omega }{\hbar v}\Theta \left(
x+vt-x_{n}\right) $.

To show the damping effect of the 1D array on the spin state of the
particles, we first study the time evolution of one particle (say, $A$)
prepared initially in the state $\left\vert \psi \right\rangle \left\vert
x_{A}\right\rangle $, where $\left\vert \psi \right\rangle =c_{0}\left\vert
\downarrow \right\rangle +c_{1}\left\vert \uparrow \right\rangle $ is the
superposition state of the spin-up $\left\vert \uparrow \right\rangle $ and
spin-down $\left\vert \downarrow \right\rangle $ and the particle is
initially located at the origin $x_{A}=0$. The initial state of the 1D $N$%
-spin array is taken to be the ground state $\left\vert 0_{E}\right\rangle $
(i.e. all spins down). The spin state of the particle at time $t>0$ reads%
\begin{eqnarray}
\notag
\rho ^{S}\left( t\right) &=&\left\vert c_{0}\right\vert
^{2}\left\vert \downarrow \right\rangle \left\langle \downarrow
\right\vert +\left\vert c_{1}\right\vert ^{2}\left\vert \uparrow
\right\rangle \left\langle \uparrow \right\vert \\
&&+c_{1}c_{0}^{\ast }\left\vert \uparrow \right\rangle \left\langle
\downarrow \right\vert f\left( t\right) +h.c.,
\end{eqnarray}%
where the off-diagonal elements are proportional to decoherence factor \cite%
{HCmodel-2}
\begin{eqnarray}
\notag
f\left( t\right) &=&\left\langle 00_{E}\right\vert
U_{A}\left\vert
00_{E}\right\rangle \\
&=&\prod\limits_{n}\cos \left[
\frac{V_{0}\Omega }{\hbar v_{A}}\Theta \left( v_{A}t-x_{n}\right)
\right]
\end{eqnarray}%
with a factorization structure. Now, we introduce the parameter%
\begin{equation}
q_{j}=\sin ^{2}\frac{V_{0}\Omega }{\hbar v_{j}},
\end{equation}%
which is the "spin-flip" probability, i.e., the probability of dissociating
one AgBr molecule due to the energy exchange with the $j$th particle. Here,
only one particle is considered i.e., $j=A$. For an array with $N$ spins, $%
\bar{n}=qN$ is the average number of dissociated molecule. In the
weak-coupling macroscopic limit%
\begin{equation}
q_{A}=\left( \frac{V_{0}\Omega }{\hbar v_{A}}\right) ^{2},
\end{equation}%
and $\bar{n}=qN$ is required to be finite \cite{HCmodel-4}. With the
assumption that
\begin{equation}
x_{n}=x_{1}+\left( n-1\right) \Delta,
\end{equation}%
the decoherence factor is approximately calculated as%
\begin{equation}
f\left( t\right) =e^{-\frac{\bar{n}}{2}\left[ \frac{v_{A}t-x_{1}}{L}\Theta
\left( x_{N}-v_{A}t\right) \Theta \left( v_{A}t-x_{1}\right) +\Theta \left(
v_{A}t-x_{N}\right) \right] }
\end{equation}%
for $\Delta /L\rightarrow 0$ as $N\rightarrow \infty $. It can be
found that the decoherence factor decays exponentially within the
regime of the macroscopic medium (i.e. $x_{1}<v_{A}t<x_{N}$), and
becomes a constant after the interaction is over. In the terminology
of quantum mechanics, the qubit is subject to a phase damping
channel.


\section{\label{Sec:4} Quantum correlations of the two qubits}


We regard the spins of the two particles as a two-qubit composite
system and the 1D $N$-spin array as the environment. To investigate
the quantum correlations of the two qubits interacting with the same
environment, we assume that the two-qubit system and the environment
are initially
uncorrelated%
\begin{equation}
\rho \left( 0\right) =\rho _{in}^{S}\otimes \rho _{in}^{D}\otimes
\rho _{in}^{E}.  \label{4eq-01}
\end{equation}%
Hereafter, the density matrixes related to the spin and spatial
degrees of the particle $Q_{A}$($Q_{B}$), and the 1D $N$-spin array
are discriminated by the superscripts $S$, $D$, $E$. To further
distinguish the spin (spatial)
degree between the particle $A$ and $B$, the superscripts $S_{\beta }$ ($%
D_{\beta }$) will be used with $\beta =A,B$. The two-qubit system is initial
in a class of states with maximally mixed marginal, known as Bell-diagonal
states \cite{xtypeLuo77}%
\begin{equation}
\rho _{in}^{S}=\frac{1}{4}\left( I_{AB}+\underset{j=1}{\overset{3}{\sum }}%
c_{j}\tau _{j}^{A}\otimes \ \tau _{j}^{B}\right),  \label{4eq-02}
\end{equation}%
which has been discussed in the literature on entanglement and its decay
under decoherence \cite{YuPRL93}, and quantum correlations besides
entanglement \cite%
{IndeEnv-1,IndeEnv-2,IndeEnv-3,IndeEnv-4,IndeEnv-5,IndeEnv-6,xtype-2}.
Here, $I_{AB}$ is the $4\times 4$ identity matrix, and $c_{j}$
$(0\leq \left\vert c_{j}\right\vert \leq 1)$ are real numbers satisfying
the unit trace and positivity conditions of the density $\rho _{in}^{S}$. The state in Eq.~(\ref%
{4eq-02}) becomes the Werner state when $\left\vert c_{1}\right\vert
=\left\vert c_{2}\right\vert =\left\vert c_{3}\right\vert =c$ and
Bell state when $\left\vert c_{1}\right\vert =\left\vert
c_{2}\right\vert =\left\vert
c_{3}\right\vert =1$. After the interaction the two-qubit state evolves into%
\begin{equation}
\rho ^{S}\left( t\right) =\frac{1}{4}\left(
\begin{array}{cccc}
1+c_{3} & 0 & 0 & \Lambda _{2}\left( t\right) \\
0 & 1-c_{3} & \Lambda _{1}^{\ast }\left( t\right) & 0 \\
0 & \Lambda _{1}\left( t\right) & 1-c_{3} & 0 \\
\Lambda _{2}^{\ast }\left( t\right) & 0 & 0 & 1+c_{3}%
\end{array}%
\right) ,  \label{4eq-03}
\end{equation}%
where the off-diagonal elements are time-dependent
\begin{subequations}
\label{4eq-04}
\begin{eqnarray}
\Lambda _{1}\left( t\right) &=&\left( c_{1}+c_{2}\right) f_{1}\left(
t\right) , \\
\Lambda _{2}\left( t\right) &=&\left( c_{1}-c_{2}\right) f_{2}\left(
t\right) ,
\end{eqnarray}%
and the diagonal elements of the density matrix $\rho ^{S}\left(
t\right) $ do not change with time. In Eq.~(\ref{4eq-04}), the
time-dependent factors are defined as
\end{subequations}
\begin{subequations}
\label{4eq-05}
\begin{eqnarray}
f_{1}\left( t\right) &=&e^{i\left( \omega _{A}-\omega _{B}\right) t}\int
dx_{A}^{\prime }dx_{B}^{\prime }\left\langle x_{A}^{\prime }x_{B}^{\prime
}\right\vert \rho _{in}^{D}\left\vert x_{A}^{\prime }x_{B}^{\prime
}\right\rangle  \notag \\
&&\text{Tr}_{E}\left[ U_{B}\left( x_{B}^{\prime },t\right) \rho
_{in}^{E}U_{A}^{\dag }\left( x_{A}^{\prime },t\right) \right] , \\
f_{2}\left( t\right) &=&e^{-i\left( \omega _{A}+\omega _{B}\right) t}\int
dx_{A}^{\prime }dx_{B}^{\prime }\left\langle x_{A}^{\prime }x_{B}^{\prime
}\right\vert \rho _{in}^{D}\left\vert x_{A}^{\prime }x_{B}^{\prime
}\right\rangle  \notag \\
&&\text{Tr}_{E}\left[ U_{A}\left( x_{A}^{\prime },t\right) U_{B}\left(
x_{B}^{\prime },t\right) \rho _{in}^{E}\right] ,
\end{eqnarray}%
where $\left\vert x_{j}^{\prime }\right\rangle $ is the eigenstate of the
coordinator operator $\hat{x}_{j}$ of the $Q_{j}$ particle.

The QD \cite{QDZerekPRL88,QDJS Zhang,CPMA Ma,CSB Su,PRA Long} is
defined as the difference between total correlations and classical
correlations in a bipartite system with the expression
$\mathscr{D}\left(\rho _{AB}\right)=\mathcal{I}\left(\rho
_{AB}\right)-\mathcal{J}\left(\rho _{AB}\right)$, where $\rho _{AB}$
is the density operator of the total system. Here, total
correlations is equal to quantum mutual information
$\mathcal{I}(\rho _{AB})=S(\rho _{A})+S(\rho _{B})-S(\rho _{AB})$,
where $S$ is the von Neumann entropy, and $\rho _{A}(\rho _{B})$ is
the reduced density matrix of the subsystem $A(B)$. Classical
correlations between the two subsystems $A$ and $B$ is the largest
information gained about one subsystem after a measurement of the
other, and it can be captured by $\mathcal{J}(\rho
_{AB})=\max\left[S(\rho _{A})-S(\rho _{A}|{\Pi_{k}^{B}})\right]$,
where $S(\rho _{A}|{\Pi_{k}^{B}})$ is the entropy of $A$ after a
measurement of $B$, and $\Pi_{k}^{B}$ is an orthogonal projective
measurement made on $B$ with $\sum_{k}\Pi_{k}^{B}=I$
\cite{xtypeLuo77}. In this paper, we choose $\Pi_{k}^{B}=|\pi_{k}\rangle%
\langle\pi_{k}|, k=1,2$, where $|\pi_{1}\rangle=\cos\theta|0\rangle+e^{i%
\phi}\sin\theta|1\rangle$, $|\pi_{2}\rangle=e^{-i\phi}\sin\theta|0\rangle-%
\cos\theta|1\rangle$, with $0\leq\theta\leq\pi/2$ and
$0\leq\phi\leq2\pi$. It is worth to stress that QD is dependent on
the subsystem on which the measurement is performed, and its
quantity is not symmetrical in general. However, in the case of
$S(\rho _{A})=S(\rho _{B})$, QD computed on measuring subsystem $A$ is equal to that on measuring subsystem $%
B$ \cite{Classical correlations}. This is an important reason why
the Bell-diagonal states are chosen as the initial states of the
two-qubit system in current literatures
\cite{IndeEnv-2,IndeEnv-4,IndeEnv-5,IndeEnv-6}.

The total correlations corresponding to the density matrix in Eq.~(\ref%
{4eq-03}) can be obtained as $\mathcal{I}(\rho ^{S}\left(
t\right))=2+\sum^{4}_{i=1}\lambda_{i}\log\lambda_{i}$, where
\end{subequations}
\begin{subequations}
\label{4eq-06}
\begin{eqnarray}
\lambda_{1,2}&=&\frac{1}{4}\left[1-c_{3}\pm|(c_{1}+c_{2})f_{1}(t)|\right], \\
\lambda_{3,4}&=&\frac{1}{4}\left[1+c_{3}\pm|(c_{1}-c_{2})f_{2}(t)|\right]
\end{eqnarray}%
are four eigenvalues of $\rho ^{S}\left( t\right)$. And the classical
correlations between the two qubits is also derived as $\mathcal{J}(\rho
^{S}\left( t\right))=\sum^{2}_{n=1}\frac{1+(-1)^{n}\chi}{2}%
\log_{2}[1+(-1)^{n}\chi]$ , where
\end{subequations}
\begin{equation}
\chi(t)=\max\left[|c_{3}|,\frac{%
|(c_{1}+c_{2})f_{1}(t)|+|(c_{1}-c_{2})f_{2}(t)|}{2}\right].  \label{4eq-07}
\end{equation}%
Therefore, QD takes the form as%
\begin{equation}
\mathscr{D}(\rho ^{S}\left(
t\right))=2+\sum^{4}_{i=1}\lambda_{i}\log\lambda_{i}-\sum^{2}_{n=1}\frac{%
1+(-1)^{n}\chi}{2}\log_{2}[1+(-1)^{n}\chi].  \label{4eq-08}
\end{equation}

Concurrence \cite{concurrence-1} quantifies the entanglement of the
state of two-qubit system $\rho_{AB}$ and is defined as $C(\rho_{AB})=\max\left\{0,%
\lambda_{1}-\lambda_{2}-\lambda_{3}-\lambda_{4}\right\}$, where $%
\lambda_{1}\geq\lambda_{2}\geq\lambda_{3}\geq\lambda_{4}$ are square roots
of the eigenvalues of the matrix $\rho_{AB}(\tau_{y}\otimes\tau_{y})%
\rho_{AB}^{\ast}(\tau_{y}\otimes\tau_{y})$, with $\rho_{AB}^{\ast}$
denoting the complex conjugate of $\rho_{AB}$ and $\tau_{y}$ being
Pauli matrix.
When the density matrix of two-qubit has an $X$ structure, defined by $%
\rho_{12}=\rho_{13}=\rho_{24}=\rho_{34}=0$, concurrence has a simple
analytic expression \cite{concurrence-2}%
\begin{equation}
C(\rho_{AB})=2\max\left\{0,\Gamma_{1},\Gamma_{2}\right\},  \label{4eq-09}
\end{equation}%
where $\Gamma_{1}=|\rho_{14}|-\sqrt{\rho_{22}\rho_{33}}$ and $%
\Gamma_{2}=|\rho_{23}|-\sqrt{\rho_{11}\rho_{44}}$. For the density operator
given by Eq.~(\ref{4eq-03}), we can get
\begin{subequations}
\label{4eq-10}
\begin{eqnarray}
\Gamma_{1} &=&\frac{1}{4}\left[ |(c_{1}-c_{2})f_{2}\left( t\right)|-|1-c_{3}|%
\right] , \\
\Gamma_{2} &=&\frac{1}{4}\left[ |(c_{1}+c_{2})f_{1}^{\ast}\left(
t\right)|-|1+c_{3}|\right].
\end{eqnarray}


\subsection{\label{Sec:4.1} Two particles located initially at the same
position}


For the sake of simplicity, we give priority to the case that the
two particles are initially located at the origin $x_{A}=x_{B}=0$,
and have the identical velocity parameter $v_{A}=v_{B}=v$. The
initial state of the macroscopic medium is taken to be the ground
state $\left\vert 0_{E}\right\rangle $. From Eq.~(\ref{4eq-05}), the
time-dependent factor is calculated as
\end{subequations}
\begin{subequations}
\label{4eq-11}
\begin{eqnarray}
f_{1}\left( t\right) &=&e^{i\left( \omega _{A}-\omega _{B}\right) t} , \\
f_{2}\left( t\right) &=&e^{-i\left( \omega _{A}+\omega _{B}\right) t}e^{-2%
\bar{n}\left[ \frac{vt-x_{1}}{L}\Theta \left( x_{N}-vt\right) \Theta \left(
vt-x_{1}\right) +\Theta \left( vt-x_{N}\right) \right]}
\end{eqnarray}%
in the weak-coupling macroscopic limit. We note that it is unnecessary to
give the value of $\omega _{A}$ and $\omega _{B}$ because QD and concurrence
only dependent on the norm of functions $f_{1}\left( t\right)$ and $%
f_{2}\left( t\right)$.

\begin{figure}[tbp]
\includegraphics[clip=true,height=5cm,width=8.5cm]{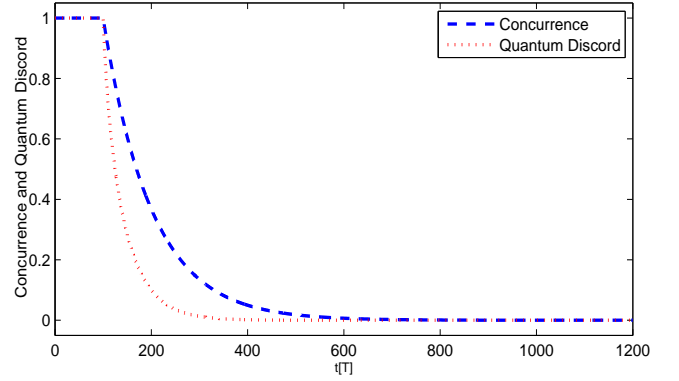}
\caption{(Color online). Dynamics of concurrence (blue-dashed line)
and
quantum discord (red-dotted line) as a function of $t$ for $c_{1}=-c_{2}=\pm1$ and $%
c_{3}=1$. Here, we choose $T=\triangle/v$ and $\triangle$ as a unit
of time and length, respectively. Moreover, $q=0.005, N=1001,
x_{1}=100, x_{N}=1100, L=1000.$} \label{fig1.eps}
\end{figure}

(1) We first consider the initial state with the following parameters: $%
c_{1}=-c_{2}=\pm 1$, $c_{3}=1$, which means the two-qubit composite
system is initially in Bell states $\left\vert \Phi ^{\pm
}\right\rangle =(\left\vert 00\right\rangle \pm \left\vert
11\right\rangle )/\sqrt{2}$. In Fig.~\ref{fig1.eps}, we plot the
concurrence $C(\rho ^{S}(t))=|f_{2}\left( t\right) |$ and the QD
$\mathscr{D}(\rho ^{S}(t))=\frac{1}{2}(1+|f_{2}\left( t\right)
|)\log (1+|f_{2}\left( t\right) |)+\frac{1}{2}(1-|f_{2}\left(
t\right) |)\log (1-|f_{2}\left( t\right) |)$ as a function of time.
It can be observed that as the time increases, the concurrence and
QD are unchanged for a while, and they decrease afterward. Actually,
the dynamics of the particles can be divided into three time
periods. The first time period ends at the particles meeting the
$1$st spin of the chain. In this time period, the state of the total
system is unchanged since the particles do not interact with the
macroscopic medium, therefore, the concurrence and QD keep their
initial value. After the particles enter the medium, the second time
period begins. Due to the energy exchange between the system and
environment, the environment results in the dephasing of the
two-qubit composite system, which decreases the concurrence and QD.
The third time period begins after the particles have interacted
with the last spin of the chain. The concurrence and QD do not
change with time due to the noninteraction between the particles and
the macroscopic medium. Here, both concurrence and QD decay
asymptotically, and the QD decays faster than concurrence.

(2) Consider the two-qubit composite system is initially in the Bell states $%
\left\vert \Psi ^{\pm }\right\rangle =(\left\vert 01\right\rangle \pm
\left\vert 10\right\rangle )/\sqrt{2}$, which is obtained by setting $%
c_{1}=c_{2}=\pm 1$ and $c_{3}=-1$ in Eq.~(\ref{4eq-02}). In this case, one
of the off-diagonal element in Eq.(\ref{4eq-03}) tends to zero, i.e., $\Lambda
_{2}\left( t\right) \rightarrow 0$. The concurrence and QD keep their initial value,
i.e., $C(\rho ^{S}(t))=\mathscr{D}(\rho ^{S}(t))=1$. The total correlations
are equally divided into classical and quantum correlations through all the
time. This phenomenon can be observed from Eq.(\ref{4eq-05}a), whose
evolution operator $U_{A}^{\dag }\left( x_{A},t\right) U_{B}\left(
x_{B},t\right) $ is generated by Hamiltonian $H_{1}^{\prime }=H_{AE}^{\prime
}-H_{BE}^{\prime }$. Since the inter-distance between two particles
vanishes, $H_{1}^{\prime }=0$, which means the environment does not induces
the loss of coherence without energy exchange. Consequently, the
off-diagonal element remains unchanged in the subspace with one spin up and
one spin down.

\begin{figure}[tbp]
\includegraphics[clip=true,height=5cm,width=8cm]{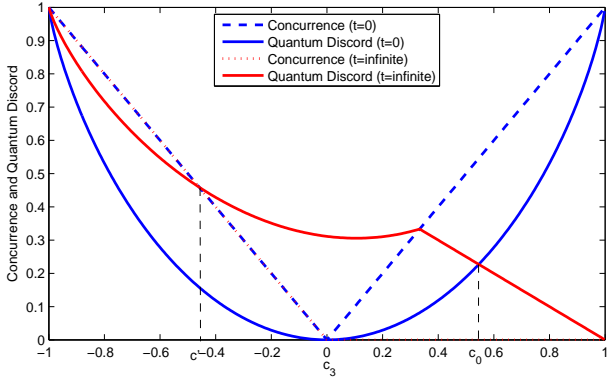}
\caption{(Color online). The concurrence and QD as a function of
$c_{3}$ at $t=0$ and $t=\infty$ with $c_{1}=\pm 1,c_{2}=\mp c_{3}$
and $|c_{3}|<1$. The other parameters are the same as those in
Fig.~\ref{fig1.eps}.} \label{fig2.eps}
\end{figure}
(3) Now, we consider the case with parameters $c_{1}=\pm 1,c_{2}=\mp
c_{3}$ and $|c_{3}|<1$ in Eq.~(\ref{4eq-02}), which is the mixture
of Bell states mentioned above~\cite{IndeEnv-6}
\end{subequations}
\begin{equation}
\rho ^{S}\left( 0\right) =\frac{1+c_{3}}{2}\left\vert \Phi ^{\pm
}\right\rangle \left\langle \Phi ^{\pm }\right\vert +\frac{1-c_{3}}{2}%
\left\vert \Psi ^{\pm }\right\rangle \left\langle \Psi ^{\pm }\right\vert .
\label{4eq-12}
\end{equation}%
To give a preliminary change of the concurrence and QD before and
after the interaction between the particles and environment, we
first discuss the quantities $C(\rho ^{S}\left( t\right) )$,
$\mathscr{D}(\rho ^{S}\left( t\right) )$ at time $t=0$ and
$t=\infty$
\begin{subequations}
\label{4eq-13}
\begin{eqnarray}
C(\rho ^{S}\left( 0\right) ) &=&|c_{3}|, \\
C(\rho ^{S}\left( \infty \right) ) &=&\left\{
\begin{array}{c}
|c_{3}|,\text{for }c_{3}<0 \\
0,\text{for }c_{3}>0%
\end{array}%
\right. \\
\mathscr{D}(\rho ^{S}\left( 0\right) ) &=&\frac{1}{2}(1-c_{3})\log
_{2}(1-c_{3})  \notag \\
&&+\frac{1}{2}(1+c_{3})\log _{2}(1+c_{3}), \\
\mathscr{D}(\rho ^{S}\left( \infty \right) ) &=&\frac{1}{2}(1-c_{3})\log
_{2}[2(1-c_{3})]  \notag \\
&&+\frac{1}{2}(1+c_{3})\log _{2}(1+c_{3})-\frac{1+\theta }{2}\log
_{2}(1+\theta )  \notag \\
&&-\frac{1-\theta }{2}\log _{2}(1-\theta ),
\end{eqnarray}%
where $\theta =\max \left[ |c_{3}|,\frac{1-c_{3}}{2}\right] $.
Equation~(\ref{4eq-13}) is obtained by letting $\Lambda_2(t)=0$ as
$\Lambda_2(t)$ approaches zero very fast. In Fig.~\ref{fig2.eps}, we
plot the concurrences and QDs at $t=0$ and $t=\infty $ as a function
of the parameter $c_{3}$. We note that the curve $C(\rho ^{S}\left(
\infty\right) )$ at $c_{3}\in(-1,0)$ overlaps the curve $C(\rho
^{S}\left(0\right) )$ in Fig.~\ref{fig2.eps}. It can be observed
that: (i) When $c_{3}\in(-1,0)$, concurrence $C(\rho ^{S}\left(
\infty\right) )=C(\rho ^{S}\left( 0\right) )$, but when
$c_{3}\in(0,1)$, $C(\rho ^{S}\left( \infty \right) )=0$, which is
always smaller than its initial value $C(\rho ^{S}\left(0\right) )$.
Actually, the variation of the concurrence before and after the
interaction is due to the increasing of the probability of state
$|\Phi^{\pm}\rangle$ as the parameter $c_{3}$ increases in
Eq.~(\ref{4eq-12}). (ii) The initial QD, $\mathscr{D}(\rho
^{S}\left( 0 \right) )$, is symmetry about the axis $c_3=0$. The
final QD, $\mathscr{D}(\rho ^{S}\left( \infty \right) )$, has a
sudden change at the point $c_{3}=1/3$, which is caused by the value
of $\theta$. When $c_{3}=-1$, the state given in Eq.~(\ref{4eq-12})
reduces to $\left\vert \Psi^{\pm }\right\rangle$, and the results
presented previously is recovered. When $c_{3}\in(-1,c_0)$, the
final QD is larger than the initial QD, i.e., $\mathscr{D}(\rho
^{S}\left(\infty \right) )>\mathscr{D}(\rho ^{S}\left(0 \right))$,
which means QD is amplified after the two-qubit composite system
interacting with the environment. As $c_{3}$ continues to increase,
$\mathscr{D}(\rho ^{S}\left(\infty \right) )<\mathscr{D}(\rho
^{S}\left(0 \right))$. (iii) In the initial time, the QD is always
smaller than concurrence, which shows that the total amount of
quantum correlation is captured by an entanglement measure. However,
things become different at $t=\infty$. Two curves intersect each
other at $c_{3}=c^{^{\prime }}$. When $c_{3}\in(-1,c^{^{\prime }})$,
the QD is smaller than the concurrence. However, the QD is greater
than the concurrence when $c_{3}\in(c^{^{\prime }},1)$, which shows
that the absence of entanglement does not necessarily indicate the
absence of quantum correlation. (iv) At the point $c_{3}=0$, there
is no correlation (QD and concurrence) between the two qubits since
the initial state in Eq.~(\ref{4eq-12}) is a maximum mix state.
After the interaction is completed, quantum entanglement remains
unchanged, however, the QD is nonzero. Hence the interaction
generates the quantum correlation.

\begin{figure*}[tbp]
\includegraphics[clip=true,height=5cm,width=16.5cm]{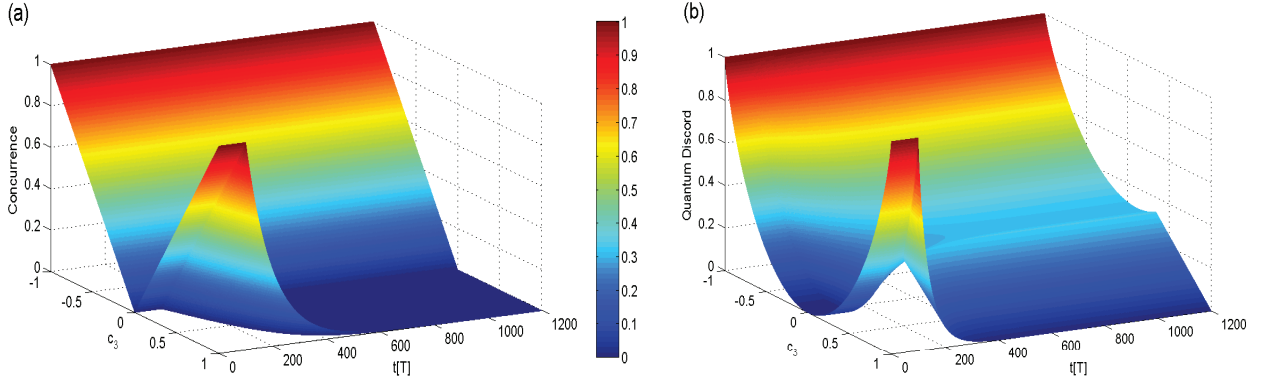}
\caption{(Color online). The dynamics of (a) concurrence and (b) QD with parameters $%
c_{1}=\pm 1,c_{2}=\mp c_{3}$ and $|c_{3}|<1$. The other parameters
are the same as those in Fig.~\protect\ref{fig1.eps}.}
\label{fig3.eps}
\end{figure*}

\begin{figure}[tbp]
\includegraphics[clip=true,height=5cm,width=8cm]{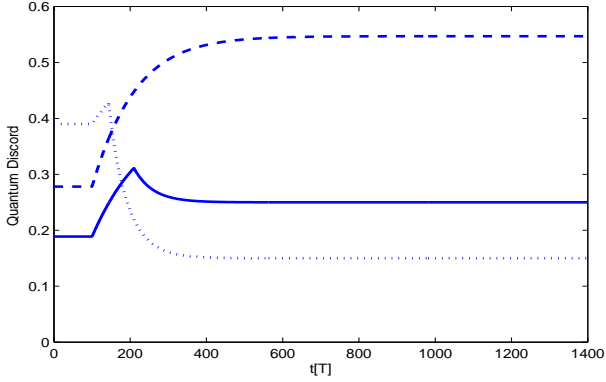}
\caption{(Color online). The dynamics of QD with parameters
$c_{1}=\pm 1,c_{2}=\mp c_{3}$ and $c_{3}=-0.6$ (dashed line), $0.5$
(solid line), $0.7$ (dotted line). The other parameters are the same
as those in Fig.~\protect\ref{fig1.eps}.} \label{fig4.eps}
\end{figure}

Now, we discuss the entanglement dynamics and the time evolution of
the QD. Actually, these are predicted in the above discussion. The
entanglement dynamics remains unchanged over the time for
$c_3\in(-1,0)$, which can be observed in Fig.~\ref{fig3.eps}(a). For
$c_3\in(0,1)$, the concurrence first keeps its initial value when
the two-qubit is away from the macroscopic medium. After the two
qubits enter the macroscopic medium, the concurrence with the
expression
$C(\rho^{S}(t))=\frac{1}{2}[|(1+c_{3})f_{2}(t)|-(1-c_{3})]$ decays
monotonically and becomes zero at time
$t_{0}=\frac{x_{1}}{v}-\frac{L}{2qNv}\ln\frac{1-c_{3}}{1+c_{3}}$.
With $c_3\in(-1,1/3)$, the QD increases monotonically over the time
after the two particles begin to interact with the macroscopic
medium, and QD gradually becomes the constant
$\mathscr{D}(\rho^{S}\left( \infty \right) )$ in Eq.(\ref{4eq-13}d),
as shown in Fig.~\ref{fig3.eps}(b) and the dashed line in
Fig.~\ref{fig4.eps}. In this regime,
$\chi(t)=\frac{|1-c_{3}|+|(1+c_{3})f_{2}(t)|}{2}$, the macroscopic
medium increases the quantum correlations between the two qubits. In
the regime $1/3<c_{3}<1$, as the interaction begins, the function
$\chi(t)=\frac{|1-c_{3}|+|(1+c_{3})f_{2}(t)|}{2}$ before a specific
time $t_c$, where
$t_{c}=\frac{x_{1}}{v}-\frac{L}{2qNv}\ln\frac{3c_{3}-1}{1+c_{3}}$.
During the period of time $t<t_c$, we observe an increasing of QD.
At time $t>t_c$, $\chi(t)=|c_{3}|$, QD decreases gradually.
Consequently, there is a sudden change in the behavior of the QD at
time $t=t_c$, as shown in the solid and dotted line in
Fig.~\ref{fig4.eps}. Actually, the increase of QD at time $t<t_c$ is
due to the classical correlation decaying faster than the total
correlation.


\subsection{\label{Sec:4.2} Two particles located initially at the
different position}


In this section, we consider that the two particles are initially located at the different
location with $x_{A}=0, x_{B}<0$, and the macroscopic medium is initially in state $|0_E\rangle$.
The time-dependent factors given in Eq.~(\ref{4eq-05}) read
\end{subequations}
\begin{subequations}
\label{4eq-14}
\begin{eqnarray}
f_{1}\left( t\right) &=&e^{i\left( \omega _{A}-\omega _{B}\right) t}e^{-%
\frac{\bar{n}}{2}\left[ \frac{vt-x_{1}}{L}\Theta \left( x_{N}-vt\right)
\Theta \left( vt-x_{1}\right) +\Theta \left( vt-x_{N}\right) \right] } \\
&&\times e^{\frac{\bar{n}}{2}\left[\frac{x_{B}+vt-x_{1}}{L}\Theta \left(
x_{N}-x_{B}-vt\right) \Theta\left( x_{B}+vt-x_{1}\right) +\Theta \left(
x_{B}+vt-x_{N}\right)\right]} , \notag\\
f_{2}\left( t\right) &=&e^{-i\left( \omega _{A}+\omega _{B}\right) t}e^{-%
\frac{\bar{n}}{2}\left[ \frac{vt-x_{1}}{L}\Theta \left(
x_{N}-vt\right) \Theta \left(
vt-x_{1}\right) +\Theta \left( vt-x_{N}\right) \right] }   \\
&&\times e^{-\frac{3\bar{n}}{2}\left[ \frac{x_{B}+vt-x_{1}}{L}\Theta \left(
x_{N}-x_{B}-vt\right) \Theta \left( x_{B}+vt-x_{1}\right) +\Theta \left(
x_{B}+vt-x_{N}\right) \right]} \notag
\end{eqnarray}
in the weak coupling macroscopic limit.

\begin{figure}[tbp]
\includegraphics[clip=true,height=5cm,width=8cm]{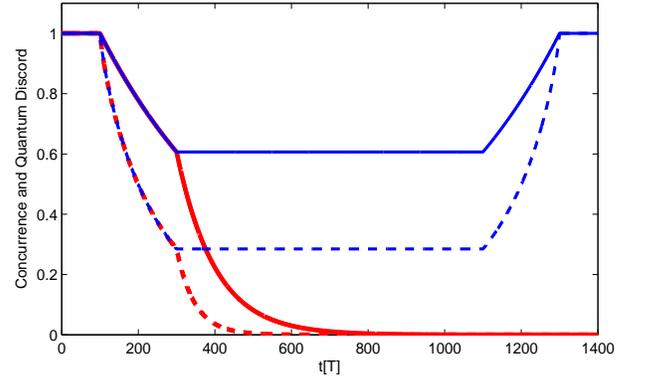}
\caption{(Color online). The dynamics of QD (dashed line) and
concurrence (solid line) as a function of $t$ for the initial state
$\Phi^{\pm}$ (red line) and $\Psi^{\pm}$ (blue line). Here, we have
set $q=0.005, N=1001, x_{1}=100, x_{N}=1100, x_{B}=-200, L=1000$.
Time is in unit of $T=\triangle/v$, and length is in unit of
$\triangle$.} \label{fig5.eps}
\end{figure}

We plot time evolution of concurrence and QD when two qubits are
initially in Bell states $\left\vert \Phi^{\pm} \right\rangle$ and
$\left\vert \Psi^{\pm}\right\rangle$ in Fig.~\ref{fig5.eps}. It can
be observed that both concurrence and quantum discord keep the
initial values before two particles meeting the macroscopic medium.
When particle $A$ begins to interact with the macroscopic medium,
they start to decrease. After the particle $B$ enters the medium,
both particles interact with the medium. However, the concurrence
and QD have different behavior for different initial states: For
initial state $\left\vert \Phi^{\pm} \right\rangle$, both
concurrence and QD decay faster than before, and vanish later; For
initial state $\left\vert \Psi^{\pm}\right\rangle$, the concurrence
and QD remain a constant when both particles interact with the
medium, and they increase after the particle $A$ left the medium and
finally reach their initial value at the time that the interaction
ends. Such process can be regarded as particle B erasing the
which-path information encoded in the medium. As the average number
of dissociated molecule $\bar{n}$ becomes larger enough, both
concurrence and QD decay to zero asymptotically as the interaction
begin, however, they were revived after a period of time for two
qubits initial in state $\left\vert \Psi^{\pm}\right\rangle$.

\begin{figure*}[tbp]
\includegraphics[clip=true,height=5cm,width=16.5cm]{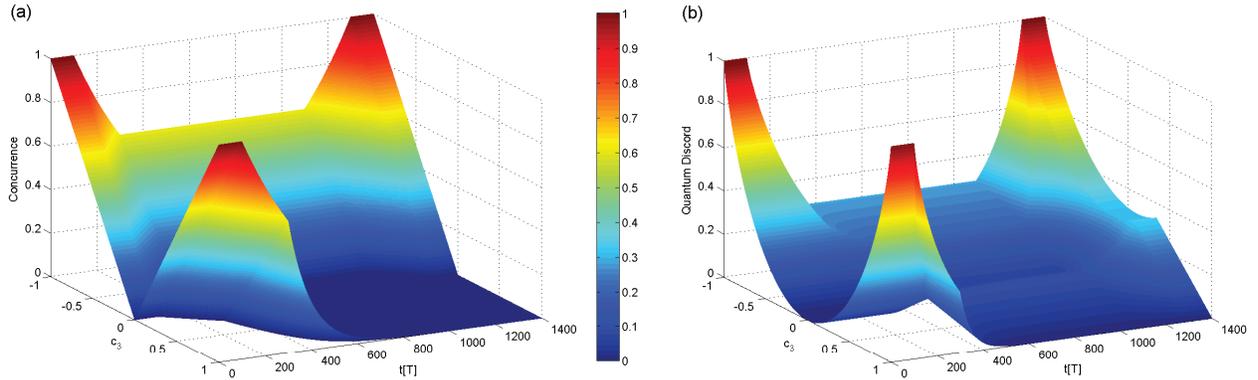}
\caption{(Color online). The dynamics of (a) concurrence and (b) QD
with parameters $c_{1}=\pm 1,c_{2}=\mp c_{3}$ and $|c_{3}|<1$. The
other parameters are the same as those in
Fig.~\protect\ref{fig5.eps}.} \label{fig6.eps}
\end{figure*}

\begin{figure}[tbp]
\includegraphics[clip=true,height=5cm,width=8cm]{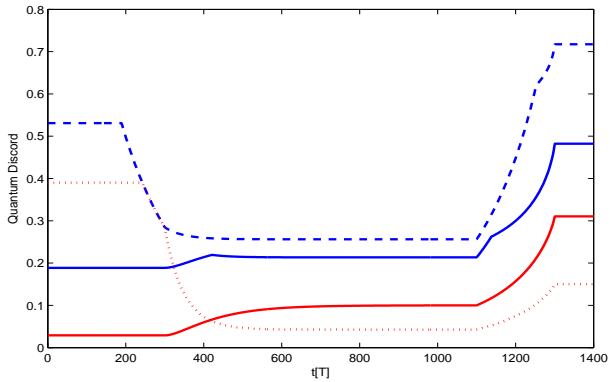}
\caption{(Color online). The dynamics of QD with parameters
$c_{1}=\pm 1,c_{2}=\mp c_{3}$ and $c_{3}=-0.8$ (blue-dashed line),
$-0.5$ (blue-solid line), $0.2$ (red-solid line), $0.7$ (red-dotted
line). The other parameters are the same as those in
Fig.~\protect\ref{fig5.eps}.} \label{fig7.eps}
\end{figure}

Now, we assume that the two-qubit system is initial in the mixture
state given in Eq.(\ref{4eq-12}). Figure~\ref{fig2.eps} also gives
the behavior of concurrence and QD before and after two particles
interacting with the medium. Consequently, the amplification and
generation of QD can also be observed in this case. To show the
difference between the previous case and the one considered here, we
plot the time evolution of concurrence and QD as the functions of
parameter $c_3$ and time in Figs.~\ref{fig6.eps}(a) and
\ref{fig6.eps}(b), respectively. It can be observed from
Fig.~\ref{fig6.eps}(a) that after the interaction begins,
entanglement first decreases, then remains unchanged for a while,
finally increases to its initial value in the regime $c_3\in(-1,0)$.
Actually, one can observed first a sudden death and latter a sudden
birth as the average number of dissociated molecule $\bar{n}$
becomes larger enough when the probability of state
$|\Psi^{\pm}\rangle$ is larger than that of state
$|\Phi^{\pm}\rangle$ in Eq.~(\ref{4eq-12}). In the regime
$c_3\in(0,1)$, only the sudden death appears.
Figs.~\ref{fig6.eps}(b) depicts that QD experiences a decline after
the interaction begins, then remains unchanged for a while, finally
revives in the case of smaller $c_3$. However, for larger $c_3$,
there is no obvious revival. Figs.~\ref{fig7.eps} gives the QD as a
function of time for a given parameter $c_3$. Here, it is can be
find that the dynamic of QD is divided into five time periods. The
first time period ends at particle $A$ meeting the $1$st spin of the
chain. In this time period, the state of two-qubit is unchanged,
then QD keeps its initial value $\mathscr{D}(\rho ^{S}\left(
0\right) )$. After particle $A$ enters the medium, the second time
period begins, and it ends at particle $B$ meeting the $1$st spin of
chain. During this period, since $f_{1}\left( t\right)=f_{2}\left(
t\right)$, QD first keeps unchanged until the critical time
$\overline{t}=\frac{x_{1}}{v}-\frac{2L\ln|c_3|}{qNv}$, then decays
monotonously, which experiences a sudden change~\cite{IndeEnv-6}.
However, such sudden change may disappears as long as $|c_3|\leq
e^{\frac{qNx_{B}}{2L}}$ as shown in the solid lines of
Fig.~\ref{fig7.eps}. The third time period is the period of both two
particles interacting with the macroscopic medium. The QD may either
increases (see the red solid line in Fig.~\ref{fig7.eps}) or
decreases (see the blue dashed line and red dotted line in
Fig.~\ref{fig7.eps}), and even has a sudden change (see the blue
solid line in Fig.~\ref{fig7.eps}), depending on the value of $c_3$.
The fourth time period begins with particle $A$ leaving the medium
and particle $B$ still interacting with the macroscopic medium. In
the case that
$\chi(t)=\frac{|(1-c_{3})f_{1}(t)|+|(1+c_{3})f_{2}(t)|}{2}$ for a
given $c_{3}$, the QD increases continuously and monotonously to the
value of $\mathscr{D}(\rho ^{S}\left( \infty \right) )$ as shown in
the red-solid line in Fig.~\ref{fig7.eps}. And in the case that
$\chi(t)=|c_{3}|$, the QD also has a continuous and monotonous
increase as shown in the red-dotted line in Fig.~\ref{fig7.eps}.
However, in the case that $\chi(t)$ is a piecewise function of time
since there exists a cross between $|c_{3}|$ and
$\frac{|(1-c_{3})f_{1}(t)|+|(1+c_{3})f_{2}(t)|}{2}$, the QD
increases fast firstly and slow later as shown in the blue-dash line
and blue-solid line in Fig.~\ref{fig7.eps}. The last time period
begins after particle $B$ have interacted with the last spin of the
chain. the QD does not change with time due to the noninteraction
between the particles and the macroscopic medium.


\section{\label{Sec:5} discussion and conclusion}


We have investigated the dynamics of concurrence and QD with an
exactly solvable model where qubits interact with a common
zero-temperature reservoir. The behaviors of both entanglement and
QD are presented in the initial states and the same reservoir
condition. It is found that regardless the inter-distance between
the two particles, the final entanglement can neither be generated
nor become larger than its initial value via the interaction.
However, QD can be amplified, and even be generated. Whether the
inter-distance of the two particles is absence or not has
significant influence on the dynamics of both entanglement and QD:
1) With a vanishing inter-distance, entanglement either remains
unchanged or decay monotonically through all the time, instead, QD
could either keep unchanged or increase monotonically, and it could
even increase first and decrease later, i.e., it undergoes a sudden
change. 2) With a nonzero inter-distance, both entanglement and QD
varied with time. when the average number of dissociated molecule
$\bar{n}$ become larger enough, entanglement experiences first a
sudden death and later a sudden birth in the time evolution, but QD
begins to decay to zero asymptotically and has a rival later.

\begin{acknowledgments}
This work was supported by NSFC Grants No. 11374095, No. 11422540, No. 11434011
and No. 11105050; NBRPC Grants No. 2012CB922103; Hunan Provincial Natural Science
Foundation of China Grants No. 11JJ7001 and No. 12JJ1002; and Scientific Research
Fund of Hunan Provincial Education Department Grant No. 11B076.
\end{acknowledgments}

\end{subequations}

\end{document}